\documentclass{IEEEtran4PSCC}
\usepackage{amsmath,amsfonts}
\usepackage{algorithmic}
\usepackage{algorithm}
\usepackage{tikz}
\usetikzlibrary{shapes.geometric, arrows}
\usepackage{textcomp}
\usepackage{diagbox}
\usepackage{xcolor}
\usepackage{graphicx}
\usepackage{algorithmic}
\tikzstyle{startstop} = [rectangle, rounded corners, minimum width=1.2cm, minimum height=0.8cm,text centered, draw=black, fill=white!20]
\tikzstyle{io} = [trapezium, trapezium left angle=70, trapezium right angle=110, minimum width=0cm, minimum height=0cm, text centered, draw=black, fill=red!30]
\tikzstyle{process} = [rectangle, minimum width=0cm, minimum height=0cm, text centered, draw=black, fill=orange!15, rounded corners]
\tikzstyle{decision} = [diamond, minimum width=1.2cm, minimum height=0.8cm, text centered, draw=black, fill=blue!10]
\tikzstyle{arrow} = [thick,->,>=stealth]
\usepackage[caption=false,font=scriptsize]{subfig}
\makeatletter
\let\old@ps@headings\ps@headings
\let\old@ps@IEEEtitlepagestyle\ps@IEEEtitlepagestyle
\def\psccfooter#1{%
    \def\ps@headings{%
        \old@ps@headings%
        \def\@oddfoot{\strut\hfill#1\hfill\strut}%
        \def\@evenfoot{\strut\hfill#1\hfill\strut}%
    }%
    \def\ps@IEEEtitlepagestyle{%
        \old@ps@IEEEtitlepagestyle%
        \def\@oddfoot{\strut\hfill#1\hfill\strut}%
        \def\@evenfoot{\strut\hfill#1\hfill\strut}%
    }%
    \ps@headings%
}
\makeatother

\psccfooter{%
        \parbox{\textwidth}{\hrulefill \\ \small{23rd Power Systems Computation Conference} \hfill \begin{minipage}{0.2\textwidth}\centering \vspace*{4pt} \includegraphics[scale=0.06]{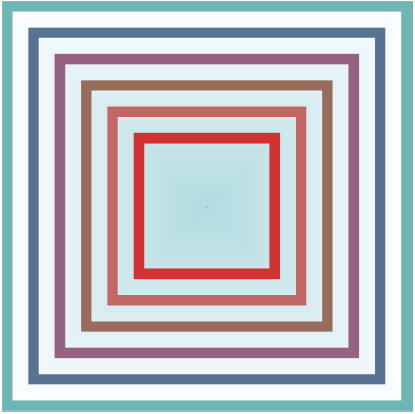}\\\small{PSCC 2024} \end{minipage} \hfill \small{Paris, France --- June 4 -- 7, 2024}}%
}

\begin{document}
\title{Optimal EV Charging Scheduling at Electric Railway Stations Under Peak Load Constraints}

%% To specify the authors when (number of affiliations <= 2)
\author{
\IEEEauthorblockN{Georgia Pierrou, Claudia Valero-De la Flor, Gabriela Hug}
\IEEEauthorblockA{Power Systems Laboratory, ETH Zurich,  Zurich, Switzerland \\ 
\{gpierrou, ghug\}@ethz.ch, cvalero@student.ethz.ch}}

%% To specify the authors when (number of affiliations > 2)
% \author{\IEEEauthorblockN{Author n.1\IEEEauthorrefmark{1},
% Author n.2\IEEEauthorrefmark{2},
% Author n.3\IEEEauthorrefmark{3}, 
% Author n.4\IEEEauthorrefmark{3} and
% Author n.5\IEEEauthorrefmark{4}}
% \IEEEauthorblockA{\IEEEauthorrefmark{1} Department Name of Organization A\\
% Name of the organization A,
% Address A\\ Emails if wanted}
% \IEEEauthorblockA{\IEEEauthorrefmark{2} Department Name of Organization B\\
% Name of the organization B,
% Address B\\ Emails if wanted}
% \IEEEauthorblockA{\IEEEauthorrefmark{3} Department Name of Organization C\\
% Name of the organization C,
% Address C\\ Emails if wanted}
% \IEEEauthorblockA{\IEEEauthorrefmark{4}Department Name of Organization D\\
% Name of the organization D,
% Address D\\ Emails if wanted}
% }

\maketitle

\begin{abstract}
In this paper, a novel Energy Management System (EMS) algorithm to achieve optimal Electric Vehicle (EV) charging scheduling at the parking lots of electric railway stations is proposed. The proposed approach uncovers the potential of leveraging EV charging flexibility to prevent overloading in the combined EV charging and railway operation along with renewable generation, railway regenerative capabilities, and energy storage. Specifically, to realize end-user flexibility, each EV state of charge at departure time is introduced as an optimization variable. Peak load constraints are included in the railway EMS to properly adjust EV charging requirements during periods of high railway demand.  A comprehensive numerical study using a scenario-tree approach on an actual railway line in Switzerland demonstrates the effectiveness and the feasibility of the proposed method in a practical setting under multiple scenarios. 
\end{abstract}


% Use this to place sponsorships
\thanksto{\noindent This work is supported in part by the ETH Mobility Initiative under MI-GRANT
2020-HS-396 and La Caixa Foundation under fellowship LCF/BQ/EU22/11930096.}

\color{black}
\vspace{-15pt}

\section*{Nomenclature}
In this section, the main nomenclature used throughout the paper is introduced. Any additional notation is defined where it first
appears in the text.
\vspace{5pt}
\addcontentsline{toc}{section}{Nomenclature}
\begin{IEEEdescription}
[\IEEEusemathlabelsep\IEEEsetlabelwidth{$V_1,V_2,V_3$}]
\item[\textit{Abbreviations}] 
\item[]
\vspace{-6pt}
\item[EMS]  Energy Management Systems
\item[ESS] Energy Storage Systems
\item[EV] Electric Vehicle
\item[PLR] Peak Load Reduction
\item[PV] Photovoltaic
\item[RB] Regenerative Braking
\item[V2G] Vehicle to Grid
\vspace{10pt}

\item[\textit{Indices}] 
\item[]
\vspace{-6pt}
\item[$t$] Time instant  
\item[$s$] Scenario

\vspace{10pt}
\item[\textit{Parameters}] 
\item[]
\vspace{-6pt}
\item[$\Delta t$] Time step
\item[$\epsilon_{B^-}$] Self-discharge coefficient
\item[$\eta_{B^+}$] ESS charging efficiency
\item[$\eta_{B^-}$] ESS discharging efficiency
\item[$\bar{P}_{B+}$] ESS charging rate
\item[$\bar{P}_{B-}$] ESS discharging rate
\item[$\eta_{EV,i}$] Electric car charging efficiency
\item[$\bar{P}_{G}$] Maximum power  drawn from the main grid
\item[$P_r$] Rated capacity of the solar installation
\item[$\bar{P}_{S}$] Maximum power sold back to the grid
\item[$\pi_s$] Probability of each scenario
\item[$r_c$]  Radiation threshold - Quadratic impact
\item[$r_{std}$]  Radiation threshold - Standard environment
\item[$SoC_{B}^{0}$] Initial ESS state of charge
\item[$SoC_{B}^{max}$] Maximum ESS state of charge
\item[$SoC_{B}^{min}$] Minimum ESS state of charge
\item[$SoC_{EV,i}^{0}$] Initial state of charge of EV \(i\)
\item[$t_{EV,i}^{a}$] Arrival time of EV \(i\)
\item[$t_{EV,i}^{d}$]  Departure time of EV \(i\)
\item[$N_{EV}$] Number of EVs
\item[$\bar{P}_{RBE}^{t,s}$]  Available power from regenerative braking
\item[$w_P,w_\theta$] Weighting factors
\item[${{P}}_{EV,i}^0$]  Nominal charging rate of EV $i$
\item[$\bar{{P}}_{EV,i}$]  Maximum charging rate of EV $i$
\item[$P_{max}$] Maximum limit for EV and railway load 
\item[$\kappa$] EV charging flexibility parameter

\vspace{10pt}

\item[\textit{Variables}] 
\item[]
\vspace{-6pt}
\item[$\beta^{t,s}$]  Measured solar radiation
\item[$C_G^{t,s}$]  Time-dependent purchasing electricity price
\item[$C_S^{t,s}$]  Time-dependent selling electricity price
\item[$P_{B^+}^{t,s}$]  ESS charging power
\item[$P_{B^-}^{t,s}$]  ESS discharging power
\item[$P_{D}^{t,s}$]  Train demand
\item[$P_{EV,i}^{t,s}$]   Charging power of EV \(i\)
\item[$P_{EV}^{t,s}$]   Aggregated EV charging power 
\item[$P_G^{t,s}$] Power bought from the main grid
\item[$P_{PV}^{t,s}$]  Active PV power output 
\item[$P_{RBE}^{t,s}$]  Regenerative braking power 
 utilized for ESS charging

\item[$P_S^{t,s}$] Power sold back to the main grid
\item[$SoC_{B}^{t,s}$] ESS state of charge
\item[$E_{EV,i}^c$]  Requested state of charge of EV $i$ upon departure
\item[$SoC_{EV,i}^{t,s}$] State of charge of EV $i$
\item[$\theta_{EV,i}^s$]  State of charge of EV $i$ upon departure
\item[$u_{B}^{t,s}$]  Binary variable to indicate ESS charging status 

\item[$u_{G}^{t,s}$] Binary variable to indicate power exchange

\end{IEEEdescription}

\color{black}

\section{Introduction}
The rapidly growing trend of transportation electrification creates new challenges in power system operation as grid load demand increases. As more private and commercial vehicles become electrified, suitable charging infrastructure and efficient Energy Management Systems (EMS) are needed to supply the required charging and reduce potential negative impacts, such as grid overloading. A promising solution lies in the integration of Electric Vehicle (EV) parking lots in electric railway systems, which have the potential to serve as major energy hubs. Such concept can leverage the available power from Regenerative Braking (RB) that is produced during the braking mode of trains as well as renewable generation and Energy Storage Systems (ESS) to satisfy the combined railway and EV charging demand \color{black}\cite{Davoodi23,Pierrou23_1, Pierrou23_2}. \color{black}

Considering the operation of EV parking lots along with electric railways has been addressed in several works in the literature. In \cite{Pierrou23_1}, an optimal EMS that minimizes the daily operating cost of a railway station with EV demand, renewable energy, ESS, and RB power is implemented. Receding horizon control is leveraged in a railway EMS in \cite{Pierrou23_2} to minimize the daily operating costs and peak power spent on EV charging. A method that fully utilizes RB power from electric railways to feed EV charging stations through a power electronics interface is proposed in \cite{Kaleybar20}. In \cite{Cicek22}, an optimal energy management strategy that utilizes the existing rail system along with RB, ESS, and renewable generation to
meet the demand for an EV parking lot is presented. A study on the capacity to charge EVs based on the energy performances of an Italian metro line and a Spanish metro line is conducted in \cite{Rodriguez19}. In \cite{Carmen11}, an energy analysis on the utilization of RB power from metro-transit systems for EV charging is realized. An optimal energy management model aiming to minimize the total line losses of the railway network while covering EV charging is proposed in \cite{Karakus23}. \color{black}Model predictive control is leveraged in \cite{Kaleybar23} to realize train-to-vehicle technology. \color{black}However, EV charging flexibility was not included.

To fully realize the potential of electric railway stations as energy hubs, it is important to take advantage of the charging coordination and flexibility of EVs. To this end, various strategies for flexible end-user EV charging scheduling aiming at different objectives, such as Peak Load Reduction (PLR) \cite{Sengor19}, may be incorporated in the EMS of electric railways with EV parking lots. Incentives, such as a charging rate reduction, can encourage EV owners choosing the ``park and rail" option to participate in a flexible charging scheme, where both technical and economic benefits can be achieved.

Limited work has been done on integrating flexible end-user EV  scheduling at electric railway systems. In \cite{Sarabi16}, a Vehicle-to-Grid (V2G) charging scheduling scheme that contributes to peak power reduction and cost savings at a railway station is presented. A method that combines V2G  for flexible charging with RB power is proposed in \cite{Golnargesi23}. Nevertheless, 
 a comprehensive solution that optimizes end-user EV flexibility and combines railway and EV charging requirements along with renewable generation, ESS, and RB power is lacking.

In this work, a novel EMS algorithm for electric railway systems considering EV charging flexibility is designed. The proposed method takes into account railway and EV charging requirements as well as renewable generation, ESS, and RB power at the railway system. To prevent grid overloading 
during periods of high railway demand, peak load constraints set by the grid operator are integrated to properly adjust EV charging requirements and realize end-user flexibility. The main contributions of the paper are as follows:
\begin{itemize}
\item An approach is proposed for the railway and EV charging operator that achieves optimal flexible EV charging scheduling, in terms of maximizing the final state of charge for the EVs participating in the flexible scheme while respecting loading limits set by the grid operator. 
 
\item The importance of utilizing various elements, such as the regenerative capabilities of trains and renewable generation is revealed, as they can be leveraged to improve flexible EV charging decisions and further maximize the final charging requirements and customer satisfaction.

\item A realistic numerical study on a railway line in Switzerland demonstrates the feasibility and effectiveness of the proposed method under various scenarios generated by a scenario-tree approach and actual historical data.

\end{itemize}

\section{Railway EMS Modeling}

\begin{figure}[!tb]%[tb]
\centering
\includegraphics[width=2.5in, keepaspectratio=true,angle=0]{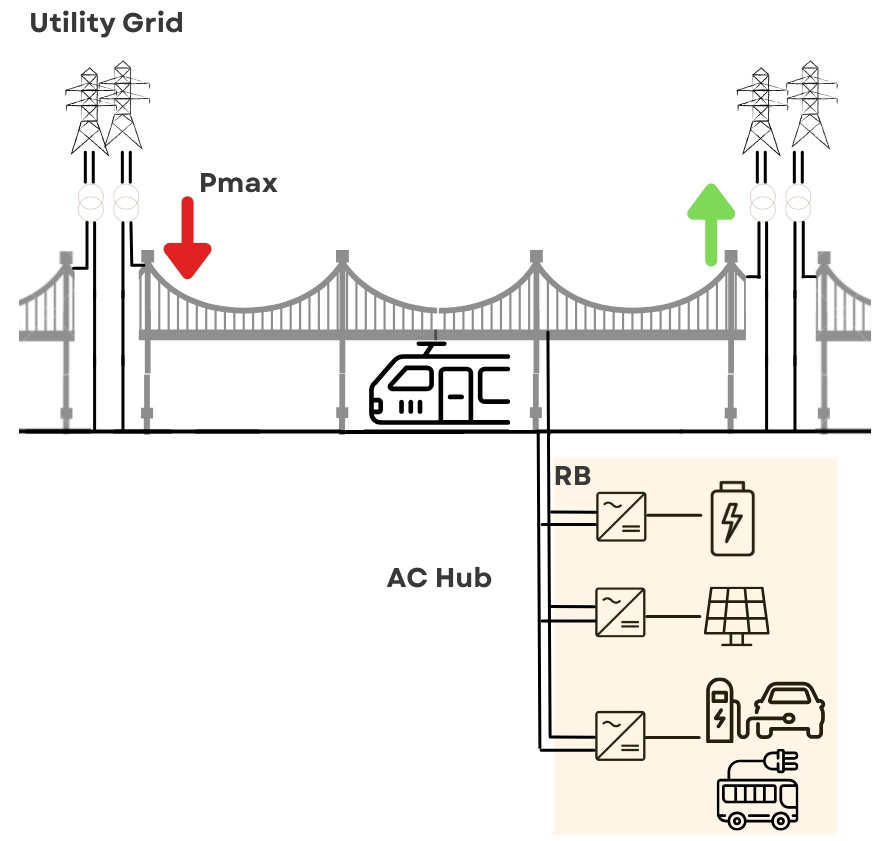}
\caption{Electric railway systems as energy hubs.}
\label{railwaystation}
\vspace{-10pt}
\end{figure}

In this work, an AC energy hub is integrated into the electric railway system, which allows the collection of electricity production from the main grid, renewable generation, or the braking of trains and its delivery to the connected loads, like trains, EV charging infrastructure, or as feedback to the grid, as illustrated in Fig. \ref{railwaystation}. Such a concept may use the mature technology of AC electric railways to fully allow the harvesting of renewable sources and supply-charging infrastructures in strategic parking areas close to railway stations, facilitating sustainable intermodal mobility \cite{magazine}. To coordinate the different types of infrastructure and quantities, such as train and EV charging demand, renewable generation, RB power, energy storage, and grid power exchange, an EMS is designed at the electric railway system level. This section describes the main components, in terms of constraints included in the railway EMS. Superscripts $s = 1, 2,..., N_s$ and $t = 1, 2, ..., N_t$
denote the considered scenario and time instant, respectively, with the time being $t \cdot \Delta t$ where $\Delta t$ is the time step size.
\subsection{Solar Generation}
Photovoltaic (PV) generation is assumed to be installed at unity power factor, i.e., only active power is considered \cite{WECC}. By utilizing historical data for the solar radiation that are available on-site, the generated active power from PVs can be estimated from the following equation relating solar radiation and the PV output \cite{Pierrouu19}:  
\vspace{-2pt}
\begin{equation}
\label{eq:solar_p_val}
P_{PV}^{t,s}={P}_{PV}^{t,s}(\beta^{t,s})=\left\{ \begin{array}{*{35}{l}}
\displaystyle \frac{{{\beta^{t,s}}^{2}}}{{{r}_{c}}{{r}_{std}}}{{P}_{r}} & 0\le\beta^{t,s}<{{r}_{c}}  \\ \\
\displaystyle \frac{\beta^{t,s}}{{{r}_{std}}}{{P}_{r}} & {{r}_{c}}\le \beta^{t,s} < {{r}_{std}}  \\ \\ 
{{P}_{r}} & \beta^{t,s} \ge {{r}_{std}}  \\
\end{array} \right.
\vspace{-2pt}
\end{equation}
\noindent where $P_{PV}^{t,s}$ is the power generated from the PV installation at the railway line, $\beta^{t,s}$ is the measured solar radiation, ${{r}_{c}}$ is a radiation threshold up to which there is a quadratic impact of the solar radiation on the PV output, ${{r}_{std}}$ is a radiation threshold in the standard environment where further increase does not lead to an increase in the PV output, and ${{P}_{r}}$ is the rated capacity of the PV installation.

% \begin{figure}[!htb]%[tb]
% \centering
% \includegraphics[width=1.5in, keepaspectratio=true,angle=0]{railway_larger (1).png}
% \caption{Electric railway stations as energy hubs.}
% \vspace{-10pt}
% \label{railwaystation}
% \end{figure}

\subsection{Power Balance}
The most important equality to be respected by the railway EMS is the balance of the electric power. Particularly, the power requirements at the train line (e.g., railway and EV charging demand, ESS charging) should be supplied by available generated power (e.g., main grid power, solar generation, ESS discharging). This translates into the following constraints that balance electric power at every time instant and scenario: 
\begin{equation}
\label{eq:powerbalance}
P_G^{t,s}+P_{PV}^{t,s}+P_{B^-}^{t,s} = P_{D}^{t,s} + P_{EV}^{t,s} + P_{B^+}^{t,s}+P_{S}^{t,s} \quad  \forall t, s
\end{equation}
where $P_G^{t,s}$ is the power supplied from the utility grid, $P_{B^-}^{t,s}$ is the power discharged by the ESS, $P_{D}^{t,s}$ is the train active power demand, $P_{EV}^{t,s}$ is the aggregated EV charging active power demand, $P_{B^+}^{t,s}$ is the power charged by the ESS, and $P_S^{t,s}$ is the power that can be sold to the main grid. 
\subsection{Grid Power Exchange}
The railway EMS model assumes that power is either bought from or sold back to the main grid. However, limitations on this grid power exchange, such as maximum power limits or the avoidance of simultaneous power exchange from and back to the grid during the same time interval should be considered. Thus, the following constraints are added to the railway EMS:
\begin{eqnarray}
\label{eq:gridlimits}
P_{G}^{t,s}&\leq& \bar{P}_{G}u_{G}^{t,s} \quad  \forall t, s \\
\label{eq:gridlimits2}
P_{S}^{t,s}&\leq& \bar{P}_{S}(1-u_{G}^{t,s})  \quad  \forall t, s \vspace{-10pt}
\end{eqnarray}
where $\bar{P}_{G}$ is a limit on the maximum power that can be bought, $\bar{P}_{S}$ is a limit on the maximum amount of power that can be sold back to the main grid, and $u_{G}^{t,s}$ is a binary variable to denote the power flow direction, i.e., it is 1 when power is bought and 0 when power is sold.
\subsection{Energy Storage Model}
A wayside ESS installed along the contact line is considered in the railway EMS. In this work, ESS may be deployed to leverage the regenerative capabilities of trains\color{black}, i.e., utilize available RB power\color{black}. Particularly, instead of being wasted or directly fed back to the main grid, available RB power is reused in the ESS charging process \cite{Pierrou23_1, Sengor18}. Although the amount of RB power that can be exploited may be limited by the ESS capacity, this approach allows to use RB power in a more flexible way, e.g., depending on EV charging needs. Hence, the ESS behavior is modeled as:
\vspace{-10pt}
\begin{eqnarray}
\label{eq:ESS1}
 P_{RBE}^{t,s} + P_{B^+}^{t,s} &\leq& \bar{P}_{B+}u_{B}^{t,s} \quad  \forall t,s \\
 \label{eq:ESS2}
P_{B^-}^{t,s}&\leq& \bar{P}_{B-}(1-u_{B}^{t,s}) \quad  \forall t,s \\
 \label{eq:ESS0}
P_{B^+}^{t,s}, P_{B^-}^{t,s}&\geq& 0 \quad  \forall t,s \\
\label{eq:ESS3}
SoC_{B}^{t,s} &=& SoC_{B}^{t-1,s} - \epsilon_{B-}SoC_{B}^{t-1,s} \\
\label{eq:ESS4}\hspace{30pt} & & + \hspace{3pt} \eta_{B+}(P_{RBE}^{t,s}+ P_{B^+}^{t,s})\Delta t \nonumber \\ & & - \hspace{3pt} \eta_{B-}P_{B^-}^{t,s}\Delta t \quad  \forall t,s \nonumber\\
\label{eq:ESS5}
SoC_{B}^{t,s} &=& SoC_{B}^{0} \quad  \forall t=t_0, s\\
\label{eq:ESS6}
SoC_{B}^{t,s}  &\leq& SoC_{B}^{max} \quad  \forall t,s\\
\label{eq:ESS7}
SoC_{B}^{t,s}  &\geq& SoC_{B}^{min} \quad  \forall t,s
\end{eqnarray}
where $P_{RBE}^{t,s}$ is the utilized RB power, $u_{B}^{t,s}$ is a binary variable to denote whether ESS is charged or discharged, i.e., it is 1 during ESS charging and 0 during ESS discharging, $SoC_{B}^{t,s}$ is the state of charge of the ESS, $\epsilon_{B-}$ is the self-discharge rate \color{black} that reflects a decrease of the shelf life of ESS\color{black}, $\eta_{B+}, \eta_{B-}$ are the charging and discharging efficiencies of the ESS, respectively, $SoC_{B}^{0}$ is the initial state of charge of the ESS, $SoC_{B}^{min}, SoC_{B}^{max}$ are the minimum and maximum levels for the state of charge of the ESS, respectively. 

\subsection{RB Power}
As shown in \eqref{eq:ESS1}, $P_{RBE}^{t,s}$ denotes the RB power that is eventually utilized for charging the ESS. However, this amount should be bounded by the available RB power. Hence, the following inequality constraints should hold:
\begin{eqnarray}
\label{eq:rbelimits}
P_{RBE}^{t,s}&\leq& \bar{P}_{RBE}^{t,s}  \quad  \forall t, s
\end{eqnarray}
where $\bar{P}_{RBE}^{t,s}$ is the available RB power extracted as negative railway demand in the overall railway demand profiles.

\section{EV Charging Modeling}
In this work, it is assumed that various types of EVs, such as electric buses or cars, can be charged at the parking lot close to the railway station. This section presents the mathematical formulation and constraints to model the individual behavior, i.e., charging power and state of charge, for each EV. 

\subsection{Aggregated EV Charging Power}
The power balance constraints formulated in \eqref{eq:powerbalance} consider the aggregated EV charging demand. This is calculated as a sum of the charging power of all plugged-in EVs as follows:
\begin{equation}
\label{eq:estimatedpower}
P_{EV}^{t,s} = \sum_{i=1}^{N_{EV}} P_{EV,i}^{t,s} \quad \forall t, s
\end{equation}
where $P_{EV,i}^{t,s}$ is the charging power of each EV, $i=1, 2, ..., N_{EV}$ and $N_{EV}$ is the number of EVs.

\subsection{Individual EV Modeling}
Regarding the state of charge of each EV, once the vehicle arrives at the parking lot, it has an initial energy level as:
\begin{equation}
\label{eq:EVSoe_initial}
SoC_{EV,i}^{t,s} = SoC_{EV,i}^{0} \quad  \forall i, s, t=t_{EV,i}^a
\end{equation}
where $SoC_{EV,i}^{t,s}$ is the state of charge at time instant $t$, $SoC_{EV,i}^{0}$ is the initial state of charge, and  $t_{EV,i}^a$ is the arrival time of EV $i$. At the time of arrival, a demanded state of charge $E_{EV,i}^c$ is also asked by the owner of EV $i$.  

Assuming that a customer has agreed to participate in a flexible charging scheme, the initially demanded state of charge may or may not be fully satisfied. However, the state of charge upon departure should respect at least some constraints to guarantee customer satisfaction to some extent. Thus, the final state of charge at the departure time of each EV is decided in a flexible range as follows: 
\begin{eqnarray}
\label{eq:finalsoc}
SoC_{EV,i}^{t,s} &\geq& \theta_{EV,i}^s  \quad  \forall i, s, t=t_{EV,i}^d \\
SoC_{EV,i}^{t,s} &\leq& E_{EV,i}^c \quad  \forall i, s, t=t_{EV,i}^d
\end{eqnarray}
where $\theta_{EV,i}^s$ is introduced as an optimization variable to realize flexibility while ensuring customer satisfaction and $t_{EV,i}^d$ is the departure time of EV $i$.

The evolution of the state of charge for each EV in charging mode is modeled using the state of charge at the previous time interval as well as the charging energy at the current time step by the following equations:
\begin{eqnarray}
\label{eq:EVSoe}
  & SoC_{EV,i}^{t,s} = SoC_{EV,i}^{t-1,s}+\eta_{EV,i} P_{EV,i}^{t,s}\Delta t & \hspace{1pt}  \\ &   \forall i, s, t \in  [t_{EV,i}^a+1, t_{EV,i}^d] &\nonumber
\end{eqnarray}
where $\eta_{EV_,i}$ is the charging efficiency.

Regarding EV charging powers, the charging power of each EV should be bounded by the maximum charging rate of the corresponding EV type as follows: 
\begin{equation}
\label{eq:maxevrate}
P_{EV,i}^{t,s} \leq \bar{P}_{EV,i} \quad  \forall i, s, t \in  [t_{EV,i}^a, t_{EV,i}^d]
\end{equation}
where $\bar{P}_{EV,i}$ is the maximum charging rate of EV $i$.

However, it is worth noting that the designated variables of each EV should only be non-zero while the vehicle is plugged in at the parking lot, i.e., only during the time period between its arrival and departure times. Therefore, the following equations should be considered:
\begin{eqnarray}
\label{eq:outofperiod}
P_{EV,i}^{t,s} &=& 0 \quad  \forall i, s, t \notin  [t_{EV,i}^a, t_{EV,i}^d]\\
SoC_{EV,i}^{t,s} &=& 0 \quad  \forall i, s, t \notin  [t_{EV,i}^a, t_{EV,i}^d]
\end{eqnarray}
% where $t_{EV,i}^a$ is the arrival time, $t_{EV,i}^d$ is the departure time, and $SoC_{EV,i}^{t,s}$ is the state of charge for the EV.

\section{The Proposed Railway EMS with Flexible EV Charging Under Peak Load Constraints}

This section presents a novel approach to optimally integrate EV charging flexibility in the EMS of electric railways under peak load constraints. Briefly speaking, peak load limitations on the combined EV charging and railway demand are included to respect technical constraints set by the utility grid regarding the transmission system, such as overloading. EV flexibility is leveraged as charging powers and the final states of charge of EVs at departure times are integrated as flexible variables to be optimized. The scheme is proposed for the railway and EV charging operator to coordinate the consumption of the flexible quantities at the railway system level, while keeping the strain added on the main grid within the physical limits required by the utility.

\subsection{Peak Load Constraints}
A limit on the maximum power to satisfy the train and EV charging demand is determined by the utility grid tasked to supply the railway line \cite{Pierrou23_2, Sengor19, Jibran21}. Hence, the following constraints are set:
\begin{equation}
\label{eq:plrlimits}
  P_{D}^{t,s} + P_{EV}^{t,s} \leq P_{max} \ \ \ \ \forall t,s
\end{equation}
where $P_{max}$ is the maximum power limit for the total railway and EV charging load at the substation where the train and EV charging stations are connected.

\subsection{EV Charging Flexibility} 
In this work, EV charging flexibility is leveraged by introducing the optimization variables  $\theta_{EV,i}^s, i=1,...,N_{EV}$ as thresholds for the EV final states of charge at departure. Equation \eqref{eq:finalsoc} has already bounded the state of charge of each EV at the departure time to the corresponding $\theta_{EV,i}^s$. Considering that departure times may differ from the fulfillment times needed to reach the initially demanded states of charge $E_{EV,i}^c$, the following upper and lower bounds are included:
\begin{eqnarray}
\label{eq:thetalimits}
\theta_{EV,i}^s&\geq& \theta_{EV,i}^{min}  \quad  \forall i, s \\
\theta_{EV,i}^s&\leq& \theta_{EV,i}^{max}  \quad  \forall i, s 
\end{eqnarray}
where \(\theta_{EV,i}^{max}\) denotes the maximum value and \(\theta_{EV,i}^{min}\) the minimum value for $\theta_{EV,i}^s$. In this paper, \(\theta_{EV,i}^{max}\) is set to be equivalent to either the initially demanded state of charge of EV \(i\) or the one achieved considering maximum charging rates in case the EV stays shorter than the fulfillment time, i.e.,:
\begin{equation}
  \theta_{EV,i}^{max} =  \min (\eta_{EV,i} \bar{{{P}}}_{EV,i} (t_{EV,i}^d-t_{EV,i}^a) \Delta t, E_{EV,i}^c)  \quad  \forall i
\end{equation}
The lower bound \(\theta_{EV,i}^{min}\) is selected as a percentage of either the initially demanded state of charge of EV \(i\) or the state of charge achieved with at least nominal charging for EV stays shorter than the fulfillment time \cite{Casini}, as follows:
\begin{equation}
\label{eq:thetamin}
  \theta_{EV,i}^{min} = \kappa \cdot  \min (\eta_{EV,i} {{P}}_{EV,i}^0 (t_{EV,i}^d-t_{EV,i}^a) \Delta t, E_{EV,i}^c)  \quad  \forall i
\end{equation}
where $\kappa$ is a flexibility parameter that determines minimum customer satisfaction upon departure and ${{P}}_{EV,i}^0$ is the nominal charging rate of EV $i$.

\subsection{Objective Function}
The objective of the proposed algorithm is two-fold: it aims to achieve economic operation by minimizing the operating costs of the railway and EV charging stations while it should also ensure EV customer satisfaction by maximizing the final state of charge of EVs at departure. Hence, the objective function is formulated as follows:
\begin{equation}
  \min \vspace{3pt} \sum_s \pi_s [w_{P}\sum_t(C_G^{t,s}P_G^{t,s}-C_S^{t,s}P_S^{t,s})\Delta t-w_{\theta}\sum_i\theta_{EV,i}^s]
  \label{eq:objectivefunction}
\end{equation}
where $\pi_s$ denotes the probability of each scenario; $C_{G}^{t,s}$ is the time-dependent purchasing electricity price; $C_{S}^{t,s}$ is the time-dependent selling electricity price; $w_P, w_{\theta} \in [0,1]$ are weighting factors to weight the relative terms \color{black} and prioritize between satisfactory EV charging or more economic operations\color{black}. It is worth noting that as the optimization problem seeks to minimize grid power while maximizing the lower bounds for the states of charge $\theta_{EV,i}^s$, the final state of charge at departure time for each EV will be equal to the corresponding $\theta_{EV,i}^s$. \color{black} In addition, the consideration of investment costs is outside the scope of the work \cite{Aguado}.\color{black}

\subsection{The Proposed Algorithm}
\label{flowchart_approach}
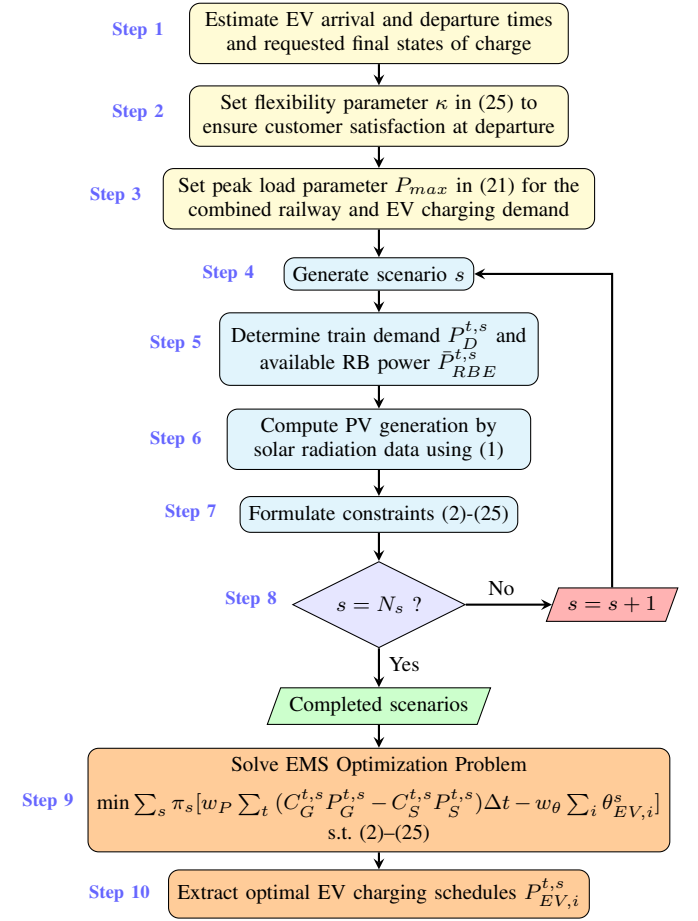
\begin{figure}[!b]
\vspace{-15pt}
\begin{center}
\begin{tikzpicture}[node distance=0.8cm]
[every label/.append style={text=red, font=\tiny}]
\tikzstyle{every node}=[font=\footnotesize]
\tikzstyle{every label}=[text=blue!60, font=\scriptsize]
\node (ev) [process, text width=4.8cm, label={[xshift=-3.2cm, yshift=-.6cm]\textbf{Step 1}}, fill=yellow!20] {Estimate EV arrival and departure times \\
and requested final states of charge};
\node (evflexibility) [process, text width=4.8cm, label={[xshift=-3.2cm, yshift=-.6cm]\textbf{Step 2}}, below of=ev, yshift=-0.3cm, fill=yellow!20] {Set flexibility parameter $\kappa$ in \eqref{eq:thetamin} to \\ensure customer satisfaction at departure};
\node (plr) [process, text width=5.5cm, label={[xshift=-3.5cm, yshift=-.6cm]\textbf{Step 3}},  below of=evflexibility, yshift=-0.3cm, fill=yellow!20] {Set peak load parameter $P_{max}$ in \eqref{eq:plrlimits} for the \\ combined railway and EV charging demand };
\node (scenarios) [process, label={[xshift=-2cm, yshift=-0.45cm]\textbf{Step 4}}, below of=plr, xshift=0cm, yshift=-0.2cm, fill=cyan!10] {Generate scenario $s$};
\node (solar) [process, text width=4cm, label={[xshift=-2.7cm,  yshift=-0.65cm]\textbf{Step 5}}, below of=scenarios, xshift=0cm, yshift=-0.2cm, fill=cyan!10] {Determine train demand $P_{D}^{t,s}$ and \\ available RB power $\bar{P}_{RBE}^{t,s}$};
\node (rbe) [process,text width=3.8cm, label={[xshift=-2.7cm,  yshift=-0.65cm]\textbf{Step 6}}, below of=solar, xshift=0cm, yshift=-0.4cm, fill=cyan!10] {Compute PV generation by \\ solar radiation data using \eqref{eq:solar_p_val}};
\node (constraints) [process, label={[xshift=-2.5cm,  yshift=-0.45cm]\textbf{Step 7}}, below of=rbe, xshift=0cm, yshift=-0.2cm, fill=cyan!10] {Formulate constraints \eqref{eq:powerbalance}-\eqref{eq:thetamin}};
\node (check_stand) [decision, label={[xshift=-1.7cm, yshift=-0.75cm]\textbf{Step 8}}, below of =constraints, yshift=-0.4cm, aspect=2] {$s=N_s$ {?}};
\node (missing) [io, right of =check_stand, xshift=2.3cm] {$s=s+1$};
\node (completed) [io, below of =check_stand, yshift = -0.55cm, fill=green!20] {Completed scenarios}; 

\node (solve) [process, text width=7.2cm, label={[xshift=-4.4cm, yshift=-0.95cm]\textbf{Step 9}}, below of=completed, text width=7.5cm, yshift=-0.45cm, fill = orange!40] {Solve EMS Optimization Problem\\ \vspace{5pt} $ \underset {} 
 {\text{min}}  \sum_s \pi_s [w_P \sum _t\hspace{0.025in} (C_{G}^{t,s}P_G^{t,s}-C_{S}^{t,s}P_S^{t,s})\Delta t-w_\theta \sum_i\theta_{EV,i}^s]$ \\ s.t. (2)--(25)};
 \node (extract) [process, label={[xshift=-3.45cm, yshift=-0.55cm]\textbf{Step 10}},below of =solve, yshift = -0.45cm, fill=orange!40] {Extract optimal EV charging schedules $P_{EV,i}^{t,s}$}; 
 \draw [arrow] (ev) -- (evflexibility);
 \draw [arrow] (evflexibility) -- (plr);
 \draw [arrow] (plr)--(scenarios);
\draw [arrow] (scenarios) -- (solar);
\draw [arrow] (solar) -- (rbe);
\draw [arrow] (rbe) -- (constraints);
\draw [arrow] (constraints) -- (check_stand);
\draw [arrow] (check_stand) -- node[anchor=west, xshift=0cm, yshift=0.05cm] {Yes} (completed);
\draw [arrow] (check_stand) -- node[anchor=south, xshift=-0.1cm, yshift=0cm] {No} (missing);
\draw [arrow] (completed) -- (solve);
\draw [arrow] (missing) |- (scenarios);
\draw [arrow] (solve)--(extract)  ;
\end{tikzpicture}
\caption{Flowchart of the proposed railway EMS with EV charging flexibility.}
\label{chart}
\end{center}
\vspace{-10pt}
\end{figure}

An illustration of the proposed algorithm is presented in Figure \ref{chart}. It can be observed that \textbf{Steps 1-3} are to determine the EV information and the peak load and flexibility parameters, whereas \textbf{Steps 4-8} are to determine the scenario-based elements, such as train demand, RB availability, and PV generation and formulate the constraints. Once all scenarios are considered and the mathematical formulation is complete, the railway EMS optimization problem is solved (\textbf{Step 9}) and flexible EV charging scheduling is extracted (\textbf{Step 10}).

\noindent \textbf{Remarks}:
\begin{itemize}
    \item Although in practice it would be suitable to solve for one scenario (e.g., a day-ahead EV charging scheduling for the parking lot run by the railway operator), multiple scenarios based on a scenario-tree approach \cite{Aguado} are considered to evaluate the impact of varying input data. Assuming that PV scenarios are $\Omega_{PV}={{P}_1, {P}_2, ..., {P_{M_1}}}$, price scenarios are $\Omega_{C_G}={C_1, C_2, ..., C_{M_2}}$, and RB power scenarios are  $\Omega_{RB}={R_1, R_2, ..., R_{M_3}}$, the final set of scenarios using the scenario-tree approach is the combination of every PV, price, and RB scenario. Thus, $\Omega_s = S_1, S_2, ..., S_{N_s}$ where $N_s = M_1 \cdot M_2 \cdot M_3$.
    \item The selection of the flexibility parameter $\kappa$ is crucial, as it determines the level of minimum customer satisfaction. Smaller values of $\kappa$ may unnecessarily decrease the EV charging levels at departure and cause dissatisfaction,
while larger values may increase EV charging requirements and cause stressed operating conditions. 

% \item The proposed method provides the decisions for the optimal flexible EV charging scheduling on a day-ahead timeframe. However, decisions can be further optimized by considering updates, e.g., on the estimated EV arrival. 
\end{itemize}

\section{Numerical Results}
This section presents the results of a comprehensive numerical study to test the proposed railway EMS integrating EV charging flexibility under peak load constraints. First, the proposed approach is validated \color{black} for one scenario \color{black} under different operation modes \color{black} to evaluate and compare the impact of various elements and inputs\color{black}, such as RB power and ESS, on the flexible EV charging decisions. Next, uncertainties associated with renewable generation, electricity prices, and RB availability are included and the results of the proposed EMS for different scenarios are presented.

\subsection{Simulation Set-up}
An actual railway line connecting Sargans and Chur in eastern Switzerland with a total length of 24.9 km and seven railway stations is used to test the proposed method. Chur is considered as the main train station where PV generation, ESS, and EV charging facilities are installed. 

Regarding data sets, daily measurements for the railway demand and RB power in 2021 as seen from the feeder directly connected to Chur's station are provided by the Swiss Federal Railways and used in the simulations. Electricity price signals consist of actual data of Switzerland's day-ahead market in 2021 \cite{ENTSOE}, where buying price $C_{G}^{t,s}$ and selling price $C_{S}^{t,s}$ are assumed to follow the same time varying trajectories. Daily solar radiation data in the area of study are also provided by the Swiss Federal Railways. A sampling time of 10 minutes is applied to avoid a high computational burden.

The daily trajectories of solar radiation are transformed to PV power using \eqref{eq:solar_p_val}. It is assumed that the solar capacity installed 
at the train station is $P_r = 1000$ kW\color{black}, which corresponds to 20$\%$ of the maximum train demand, \color{black} and the radiation parameters are set as $r_c=150$ $\text{W/m}^2$ and $r_{std}=1000$ $\text{W/m}^\text{2}.$

The ESS capacity installed is assumed as $SoC_B^{max}=1000$ kWh \color{black}\cite{Lanz19}\color{black}, whereas the minimum level for the ESS $SoC_B^{min}$ is $10\%$ of the total capacity. The ESS charging and discharging rates are set as $1000$ kW/min and the charging and discharging efficiencies are $\eta_{B+}=\eta_{B-}=0.95$. The initial energy level for the ESS $SoC_B^{0}$ is set as $50\%$ of the ESS capacity. No self-discharge phenomenon is considered, i.e., $\epsilon_{B-}=0$.

Regarding EV charging requirements, the EV parking lot at the railway station may serve different types of vehicles, such as public electric buses and private electric cars. The charging facilities are open to EV arrivals daily from 6:00 to 22:00, whereas departure can happen at any time. \color{black}The total number of EVs considered is $N_{EV}=179$. \color{black}To fully leverage the flexibility potential of EVs, it is assumed that all vehicles entering the parking lot participate in the flexible charging scheme under peak load constraints. To ensure customer satisfaction, the parameter $\kappa$ is set as 0.6 in \eqref{eq:thetamin}, meaning that upon departure all EVs will be charged to at least 60$\%$ of either their initially requested state of charge or the equivalent one if nominal charging was applied for EVs with shorter parking stays. The peak load limit set by the utility grid for the combined railway and EV charging demand throughout the day is $3000$ kW, i.e., $P_{max}=3000$ kW in \eqref{eq:plrlimits}.

EV arrival for electric cars is simulated using a probabilistic approach. The arrival times of private EVs are obtained by an exponential distribution with a rate of four, i.e., four electric cars are expected on average every hour. The departure times of private electric cars are determined using a triangular distribution, within a range of two hours with respect to the fulfillment time required for achieving the demanded state of charge levels. The initially demanded states of charge are obtained from a uniform distribution in the interval [10, 50] kWh. Nominal and maximum charging power rates for electric cars are $P_{EV,i}^0=11$ kW and $\bar{P}_{EV,i}=22$ kW, respectively \color{black}\cite{Casini,Casini2}\color{black}.
 
EV arrival for electric buses is simulated according to the official timetable for the main bus stop closest to Chur's train station. Departure times are fixed as posted on the public schedule \cite{postauto}  whereas arrival times are obtained from a triangular distribution, allowing for a range of 10-60 minutes prior to the departure time. The initially requested states of charge are obtained from a uniform distribution in the interval [100, 300] kWh. Nominal and maximum charging power rates for electric buses are $P_{EV,i}^0=\bar{P}_{EV,i}=300$ kW.

The proposed railway EMS is simulated in a \textsc{Matlab}$^\copyright$ environment. The commercially available linear programming solver Gurobi \cite{gurobi} is deployed to solve the optimization problem and obtain the optimal EV charging scheduling decisions in $\textbf{Step 9}$ of the proposed algorithm. \color{black} The weighting factors are selected as $w_P= w_{\theta}=1$. \color{black}

\subsection{Validation of the Proposed Railway EMS}
\begin{figure}[!tb]
\vspace{-12pt}
\centering
\subfloat[$\text{Train demand}$]{\includegraphics[width=1.695in]{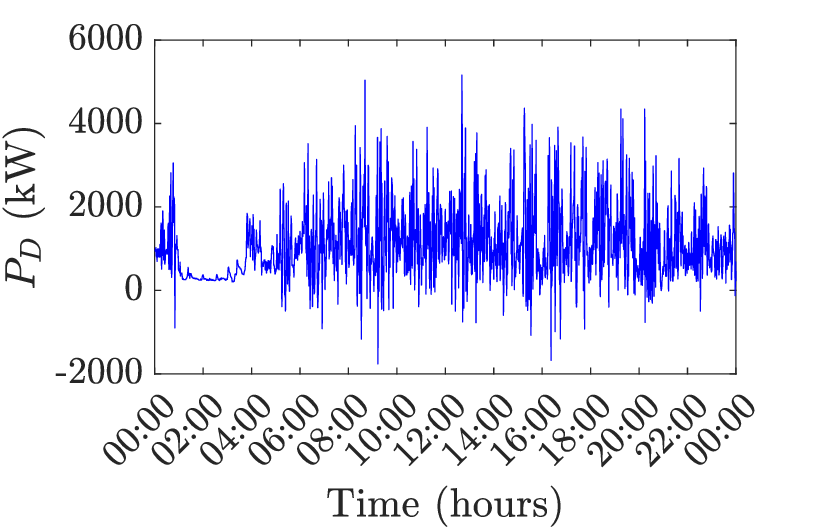}
\label{demand1}}
\hfil
\subfloat[RB power]{\includegraphics[width=1.695in]{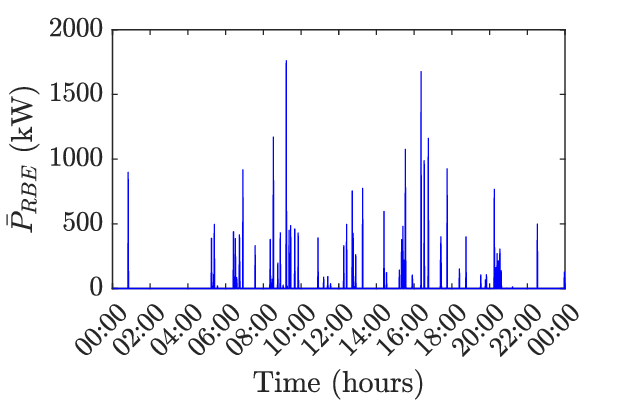}
\label{demand2}}
\vspace{5pt}
\caption{Railway power consumption and RB power for the selected scenario.} \label{railwayrbpower}
\subfloat[PV generation]{\includegraphics[width=1.665in]{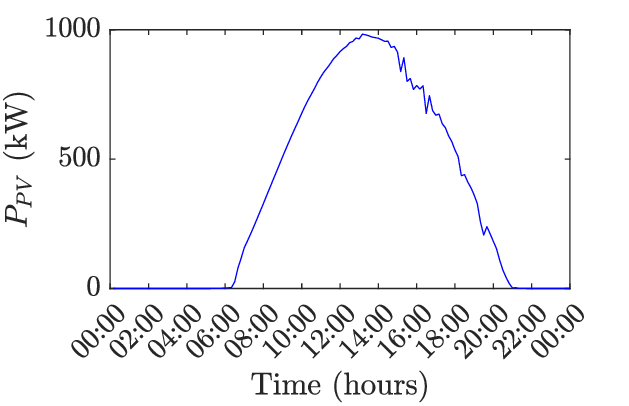}
\label{rbe1}}
\hfil
\subfloat[Day-ahead price]{\includegraphics[width=1.665in]{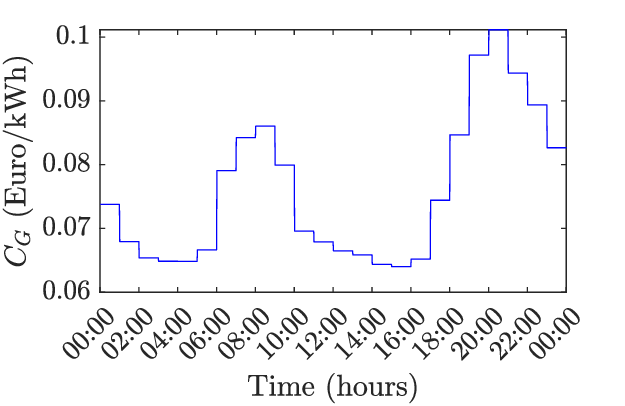}
\label{price2}}
\vspace{5pt}
\caption{PV generation and day-ahead electricity price for the selected scenario.}
\label{pvprice}
\vspace{-15pt}
\end{figure}

To validate the proposed approach and evaluate the impact of different elements on its performance, different cases are analyzed. For demonstration, one scenario that corresponds to the weekday with the highest amount of PV generation is selected. \color{black}Running one scenario takes 3.944102 seconds on a 12th Gen Intel(R) Core(TM) i7-1260P 2.10 GHz. \color{black} Figures \ref{railwayrbpower} and \ref{pvprice} show the main inputs, including the daily train demand, PV generation, RB power, and electricity price, for the selected scenario.

\begin{figure*}[!tbp]
\centering
\includegraphics[width=16cm]{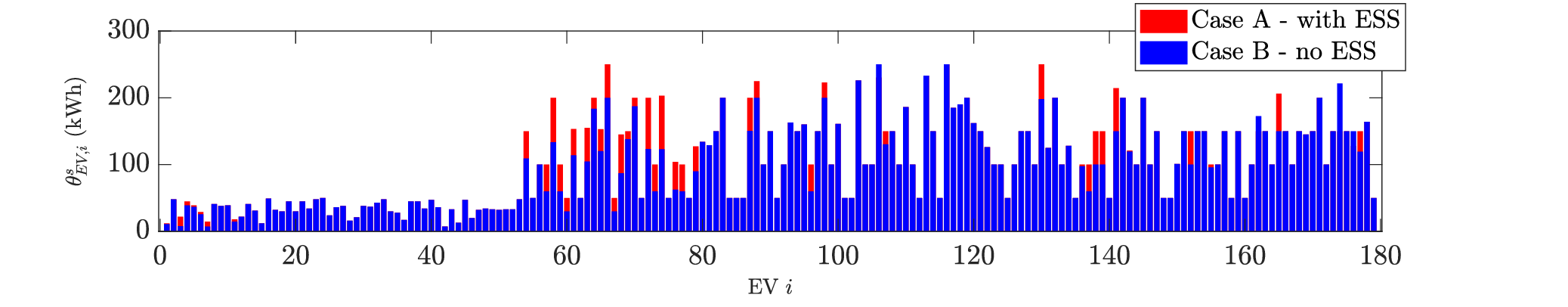}
\vspace{-10pt}
\caption{The impact of RB power on the flexible EV state of charge at departure time.}
\label{fig:theta-avsb}
\vspace{-5pt}
\end{figure*}
\begin{figure*}[!tbp]
\centering
\includegraphics[width=16cm]{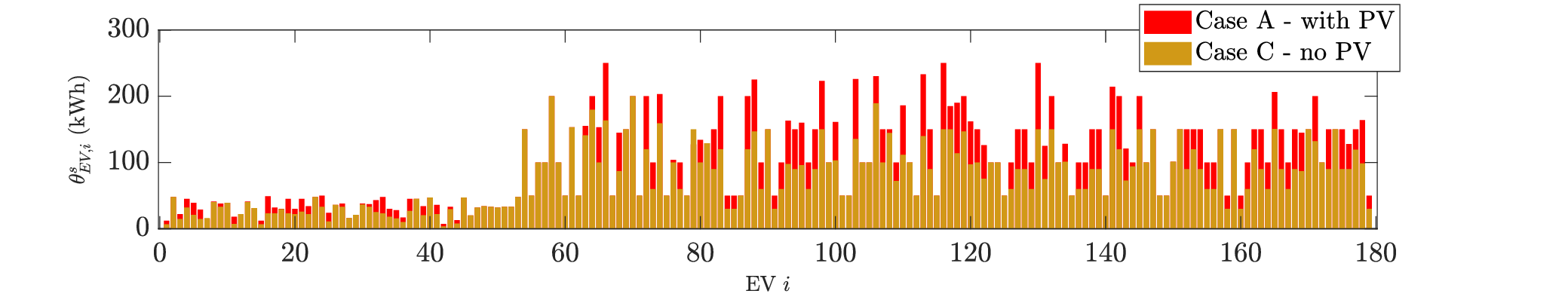}
\vspace{-10pt}
\caption{The impact of PV generation on the flexible EV state of charge at departure time.}
\label{fig:theta-avsc}
\vspace{-10pt}
\end{figure*}
% \begin{figure*}[!tbp]
% \centering
% \includegraphics[width=\textwidth]{caseavscasec_wider_grey.eps}
% \vspace{-10pt}
% \caption{The impact of PV generation on the flexible EV state of charge at departure time.}
% \label{fig:theta-avsc}
% \vspace{-5pt}
% \end{figure*}

\subsubsection{Effect of RB Power}
First, we study the performance of the proposed method with respect to the utilization of the regenerative braking capabilities of trains. The following two different cases are implemented and compared:
\begin{itemize}
 \item \textit{Case A (with ESS)}: RB power and ESS are leveraged and included in the formulation, i.e., constraints \eqref{eq:ESS1}--\eqref{eq:rbelimits} are considered.
    \item \textit{Case B (no ESS)}: RB power and ESS are not included in the formulation, \color{black}i.e., there is no ESS to store the RB power and constraints \eqref{eq:ESS1}--\eqref{eq:rbelimits} are neglected. Any power from the regenerative capabilities of trains is dissipated as waste.\color{black} 
   
\end{itemize}
The results of the final state of charge of each EV at the departure time for Cases A and B and the selected scenario are presented in Fig. \ref{fig:theta-avsb}. By comparing the two cases, the advantage of utilizing RB power in the combined railway and EV charging is highlighted. It can be observed that \textit{Case A} achieves better EV charging results, ensuring more effective customer satisfaction. Indeed, due to the utilization of RB power and ESS, an increase of up to $39.43\%$ in the final state of charge can be achieved, which corresponds to vehicle 74. In addition, the impact of RB power and ESS is more beneficial to the flexible EV charging decisions for public electric buses rather than private electric cars. This conclusion aligns with the theoretical expectation that EVs with larger capacities have greater potential for flexibility.

\subsubsection{Effect of PV Generation}
After showing the potential benefits of the utilization of RB power and ESS on the combined railway and flexible EV charging operation, the impact of PV generation is examined. The following cases are analyzed:
\begin{itemize}
 \item \textit{Case A (with PV, with ESS)}: same as before, i.e., RB power and PV generation are considered.
    \item \textit{Case C (no PV, with ESS)}: The availability of PV generation is not considered, i.e., $P_{PV}^{t,s}$ is assumed to be 0.
\end{itemize}
The resulting states of charge for EVs during departure for Cases A and C and the selected scenario are shown in Fig. \ref{fig:theta-avsc}. It can be concluded that available PV generation may greatly contribute to further maximizing the final EV charging result. In fact, by utilizing the available solar generation, an increase of up to $40\%$ in the final state of charge can be achieved, which corresponds to vehicle 116. Hence, it is important to fully leverage the available elements, such as PV and ESS, in the railway and EV charging station operation.

\subsection{A Scenario-Based Study}
To further evaluate the performance of the proposed algorithm under variations in input data, multiple scenarios are considered. Specifically, $N_s=100$ scenarios of equal probability $\pi_s=0.01$ are generated with the scenario-tree approach described in Section \ref{flowchart_approach}. The generated scenarios are based on all possible combinations of 4 PV scenarios, 5 price scenarios, and 5 RB power scenarios. Only electric buses are considered in the scenario-based study to highlight the effectiveness of the proposed approach for larger EV charging requests while reducing the overall computational burden. 

\begin{figure}[!tb]
\centering
\includegraphics[width=3.9in, keepaspectratio=true,angle=0]{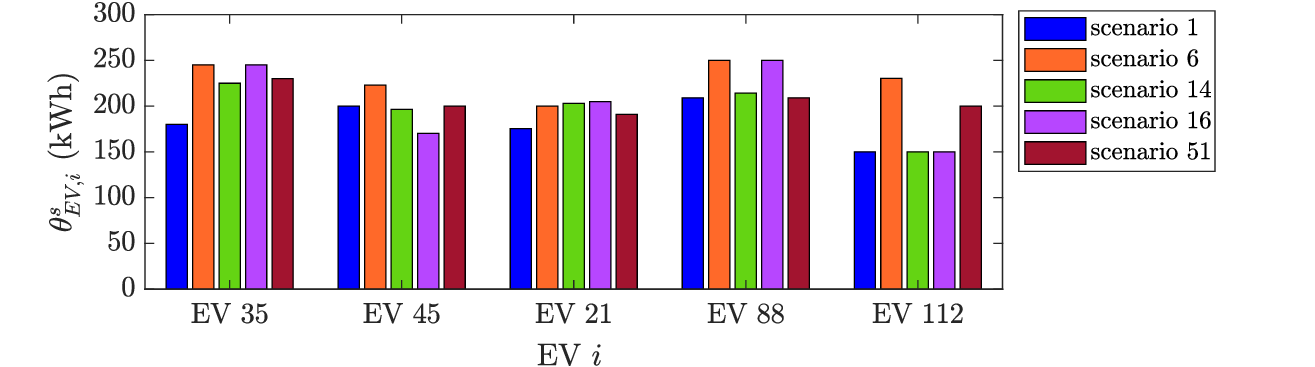}
\caption{The state of charge at departure time for selected EVs and scenarios.}
\label{scenariosmany}
\vspace{-5pt}
\end{figure}

\begin{figure}[!tb]%[tb]
\vspace{-5pt}
\centering
\includegraphics[width=3in, keepaspectratio=true,angle=0]{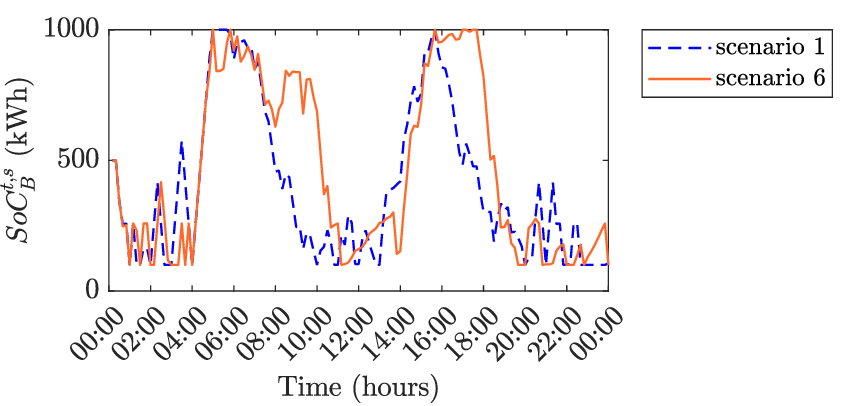}
\caption{The ESS state of charge for selected scenarios.}
\label{scenariosmany_1}
\vspace{-10pt}
\end{figure}

\begin{figure}[!tb]%[tb]
\centering
\includegraphics[width=3.2in, keepaspectratio=true,angle=0]{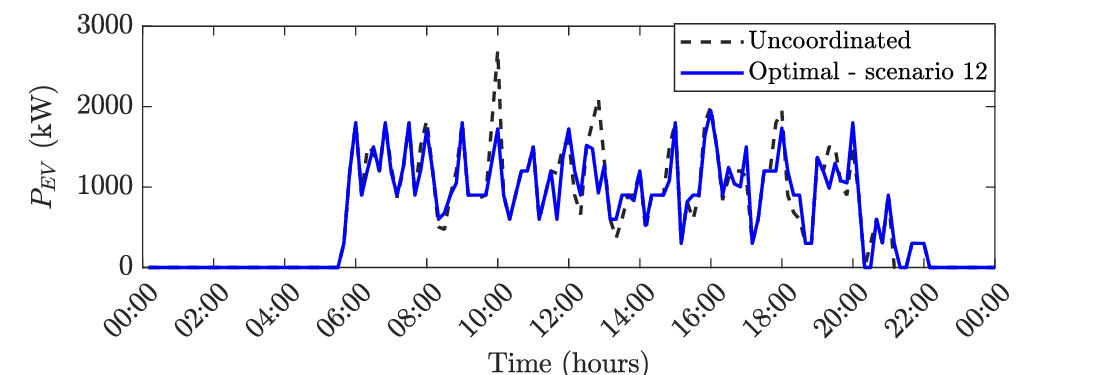}
\caption{Optimized and uncoordinated power peaks.}
\label{scenario12}
\vspace{-15pt}
\end{figure}

The results of the analysis of multiple scenarios are presented in Figs. \ref{scenariosmany}-\ref{scenario12}. In Fig. \ref{scenariosmany}, the states of charge at the departure times for selected EVs and \color{black}selected \color{black} scenarios \color{black}of different solar generation, price, and RB power \color{black}are presented. It can be seen that the proposed railway EMS effectively adjusts individual EV charging requirements with respect to the variations of inputs. For instance, considering EV $112$, the optimal state of charge at departure time may greatly vary depending on the scenario (e.g., 150 kWh in scenarios 1, 14, and 16, 230 kWh in scenario 6, and 200 kWh in scenario 51). This behavior can be well explained by looking at the differences between the illustrated scenarios. Indeed, the results of scenario $1$ are less satisfactory compared to other scenarios for the selected EVs, as it is characterized by a significantly smaller PV generation as well as lower RB power availability. Hence, as expected, other scenarios offer more satisfactory charging results for most EVs. Fig. \ref{scenariosmany_1} illustrates the ESS energy level for scenarios 1 and 6, as the scenarios with the least and most satisfactory performances, respectively. It can be observed that ESS activation is more frequent in scenario 6 due to larger RB availability. 

In addition, Fig. \ref{scenario12} illustrates a comparison of the aggregated EV charging power obtained by the proposed approach in one scenario with the one obtained by uncoordinated charging with nominal charging power. Thanks to the integration of peak load constraints, the proposed method can effectively reduce the power spikes observed in the uncoordinated charging case. Other scenarios exhibit similar behavior. Therefore, it can be concluded that the proposed algorithm can effectively leverage EV charging flexibility under multiple scenarios while satisfying customer requirements and avoiding overloading.

\section{Conclusion}
A novel EMS algorithm for optimal EV charging scheduling in the combined electric railway and EV charging operation is proposed. The proposed approach leverages EV flexibility in an effort to respect technical constraints, such as overloading. Particularly, EV charging requirements are considered as flexible variables and peak load limitations set by the main grid operator are included for the aggregated EV charging and railway demand. The proposed method represents the first attempt to uncover the potential of utilizing EV flexibility while considering renewable generation, RB power, and ESS at the electric railway station level. Numerical results validate the importance of different elements, such as PV and RB power, as well as the effectiveness under various scenarios in a practical setting. Future work may focus on including the impact of \color{black}EV battery degradation as well as \color{black}updates in the prediction of uncertainties and EV arrival.
% An example is shown in Table~\ref{table_example}.
% \begin{table}[!ht]
% % increase table row spacing, adjust to taste
% \renewcommand{\arraystretch}{1.3}
% \centering
% \caption{This is a table.}
% \label{table_example}
% \begin{tabular}{|c|c|}
% \hline
% One & Two\\
% \hline
% Three & Four\\
% \hline
% \end{tabular}
% \end{table}

\section*{Acknowledgments}
The authors would like to thank Robert Strietzel and the Swiss Federal Railways for providing the railway consumption and solar radiation data and for interesting industry insights.

\end{document}